\documentclass{aa}  
\bibpunct{(}{)}{;}{a}{}{,} % to follow the A&A style
\usepackage{graphicx}
\usepackage{txfonts}
\begin{document}

   \title{The Origin of Fluorine: Abundances in AGB Carbon Stars Revisited}

%   \subtitle{}

   \author{C. Abia\inst{1}
          \and
           K. Cunha\inst{2}
          \and
           S. Cristallo\inst{3,4}
          \and
           P. de Laverny\inst{5}
          }

\institute{Dpto. F\'\i sica Te\'orica y del Cosmos, Universidad de Granada, 18071 Granada, Spain
           \email{cabia@ugr.es}
         \and
         Observat\'orio Nacional, Rua General Jos\'e Cristino, 77, 20921-400 S\~ao Critov\~ao, 
         Rio de Janeiro, RJ, Brazil
         \and
         INAF, Osservatorio Astronomico di Collurania, 64100 Teramo, Italy
         \and
         INFN Sezione Napoli, Napoli, Italy
         \and 
         Laboratoire Lagrange, Universit\'e C\^ote d'Azur, Observatoire de la 
         C\^ote d'Azur, CNRS, CS 34229, 06304 Nice cedex 4, France
    }

   \date{Received ; accepted }
\authorrunning{C. Abia et al.}
\titlerunning{The origin of fluorine}
% \abstract{}{}{}{}{} 
% 5 {} token are mandatory
 
  \abstract
  % context heading (optional)
  % {} leave it empty if necessary  
   {Revised spectroscopic parameters for the HF molecule and a new CN line list in the 2.3 $\mu$m 
region have been recently available, allowing a revision of the F content in AGB stars.}
  % aims heading (mandatory)
   {AGB carbon stars are the only observationally confirmed sources of fluorine. Nowadays there 
is not a consensus
on the relevance of AGB stars in its Galactic chemical evolution. The aim of this 
article is to better constrain
the contribution of these stars with a more accurate estimate of their fluorine abundances.} 
% methods heading (mandatory)
   {Using new spectroscopic tools and LTE spectral synthesis, we redetermine fluorine abundances from several HF lines in 
the K-band in a sample of Galactic and extragalactic AGB carbon stars  of spectral types N, J and SC 
spanning a wide range of metallicities. }
  % results heading (mandatory)
   {On average, the new derived fluorine abundances are systematically lower by 0.33 dex with respect
to previous determinations. This may derive from a combination of the lower excitation energies of the HF lines and the 
larger macroturbulence
parameters used here as well as from the new adopted CN line list.
Yet, theoretical nucleosynthesis models in AGB stars agree with the new fluorine determinations at solar metallicities. 
At low metallicities, an agreement between theory and observations can be found by handling in a different way 
the radiative/convective interface at the base of the convective envelope.}
  % conclusions heading (optional), leave it empty if necessary 
   {New fluorine spectroscopic measurements agree with theoretical models at low and at solar metallicity. Despite this, 
complementary sources are needed to explain its observed abundance in the solar neighbourhood. }


   \maketitle
%
%________________________________________________________________

\section{Introduction}

 Fluorine has only one stable  isotope, $^{19}$F, yet fragile and easily
 destroyed  in stellar  interiors by  proton, neutron  and alpha  particle
 capture    reactions. Major fluorine destruction channels are $^{19}$F$(p,\alpha)^{16}$O    and
 $^{19}$F$(\alpha,p)^{22}$Ne reactions. Thus, any  production mechanism has also
 to  enable $^{19}$F to  escape from  the  hot stellar  interiors after  its
 production.  This makes  fluorine a  useful tracer  of  the physical
 conditions  prevailing in stellar  interiors. Fluorine  is  thought to  be  produced in  several
 scenarios: (i) Gravitational  supernovae through the  neutrino spallation
 process, (ii) low and intermediate mass (M$\leq 7$ M$_\odot$) Asymptotic  Giant Branch  (AGB) 
stars\footnote{{\bf In the following, we define Low Mass Stars (LMS) to be those with initial masses
between about 0.8 and 2.5 M$_\odot$ which experience He ignition under degenerate conditions. Stars
more massive than this succeed in igniting He gently but they are not sufficiently massive to ignite
C in their core. We named these Intermediate Mass Stars (IMS) having initial masses about 2.5-7 M$_\odot$.}} 
during  the He-burning
 Thermal  Pulses  (TP) and the subsequent Third Dredge Up (TDU) episodes,  (iii) Wolf-Rayet  (WR)  stars  
in  the  hydrostatic
 He-burning  phase  and,  even, (iv) during  white  dwarfs  mergers  \citep[see
   e.g.][]{woo90,for92,mey00,lon11}. To date, proofs of their production have been found in AGB stars 
only \citep[e.g.][]{jor92,wer05,ots08,abi10,luc11}.  The  origin  of  this
 element is therefore still  rather uncertain and contributions from
 the  above mentioned  sources  seem to  be  necessary when  comparing
 Galactic     Chemical    Evolution     (GCE)     model    predictions
 \citep{ren04,kob11}  to the evolution  inferred
from observations in Galactic field dwarf and  giant stars \citep{ale12,jon14a}.

To determine  the relative  role of the above proposed
sources   of  fluorine,   in   particular  of   AGB  stars,   accurate
determinations  of F  abundances are needed.  Fluorine
abundances  are usually  determined  in  cool dwarfs  and  giants  from
ro-vibrational  HF   lines  at  2.3  $\mu$m.   This  spectral  region,
unfortunately, is placed  in a region with a lot  of telluric lines that can be difficult
to be removed. This prevents, in many cases, the use of  several HF
lines to determine  the F abundance in a given star.  The most used
one is the  R9 $\lambda~2.3358~\mu$m  HF line, which  is marginally affected   by  blends   and   telluric  lines
\citep{abi09,lav13}.  \citet{jor92}  firstly used  this  and other  secondary HF
lines  to  derive F  abundances  in AGB  stars of both oxygen-  and
carbon-rich  types\footnote{AGB  carbon  stars  are those  with
surface C/O$>1$ (by number).} showing  indeed that AGB
stars  present  large fluorine   enhancements.   Most   recently,
\citet{abi09,abi10}  revisited this  pioneering analysis  using  a new
grid of  atmosphere models  for AGB carbon  stars  and an
up-dated  molecular   and  atomic  line   lists  in  the   2.3  $\mu$m
region. Yet, they found  significant F enhancements but systematically
lower   than  those   in  \citet{jor92}.   Nevertheless,  the   new  F
enhancements  are  in  a   global  agreement  with  
nucleosynthesis calculations  of low mass AGB
stars  with near  solar  metallicity \citep{cri11},  although at  lower
metallicities  models   predict  larger  enhancements   than  observed
\citep{abi11,luc11}.  The mentioned abundance  analyses in  AGB stars,
however, were performed using HF line excitation energies inconsistent
with  the  partition  functions  adopted  in  the  spectral  synthesis.  
In  fact, \citet{abi09,abi10,abi11}  used  the  $Turbospectrum~v10.2$  radiative transfer code \citep{ple12},  which  adopts
partition   functions  from  \citet{sau84}.   However,  the   HF  line
excitation energies  used in those  works were from  \citet{jor92} (in
turn from Tipping, private  communication), which are $0.25$ eV higher
than  the  correct  values  if  using  the  \citet{sau84}'s  partition
functions.  As mentioned  by  \citet{jon14b}, the  0.25 eV  different
adopted excitation  energies  comes  from  the fact that  the
Tipping-list uses the dissociation energy of the energy potential, and
not,  like \citet{sau84},  the  true energy  required  for dissociation.  The
former is higher due to the zero  point of the energy
of the lowest vibrational level. The difference is exactly 0.25 eV for
HF \citep{Zem91}. Recently, \citet{jon14b} presented a complete  and comprehensive 
calculation of  the  excitation  energies  and  transition  probabilities  for the HF
molecule in the K- and L-bands. The computed HF  molecular parameters agree
nicely  with  the  new   list  of  experimental  molecular  parameters
delivered  by  the HITRAN  database  \citep[see][for  details on  this
database]{rot13}. As a consequence and  according to the simple curve  of growth, F  abundances
derived with consistent partition functions    would   be   lower   by   an   amount
$\sim\theta\Delta\chi$,      where      $\theta=5040/T_{eff}$      and
$\Delta\chi=0.25$ eV.

Very recently, a new solar F abundance has been determined from the analysis of a sunspot spectrum \citep{mai14}. 
The new solar abundance, log $\epsilon$(F)$=4.40\pm 0.25$\footnote{The abundances are given using the usual definition
log $\epsilon$(X)$= 12 +$ log (X/H), where (X/H) is the abundance of the element X by number 
and log $\epsilon$(H)$\equiv 12$.}, is significantly lower that the previous commonly
adopted value $4.56\pm0.30$ although in agreement within the error bars \citep{aps09}, and it is in a very good agreement with the meteoritic 
value, $4.46\pm0.06$ \citep{lod09}. Finally, a new CN line list has been provided
that might affect the fluorine abundance determination in carbon-rich stars from HF lines in the 2.3 $\mu$m 
region \citep{hed13}. As a consequence, in  light of the new molecular parameters 
and solar fluorine abundance, a re-determination of fluorine
abundances in AGB stars is needed to evaluate their actual ontribution to the 
Galactic fluorine inventory. This is the aim of the present work.

In Section 2, we describe the observations and the new spectroscopic tools
used for the re-determination of the fluorine abundances in AGB carbon stars.
Section 3 shows our results, while we present our conclusions in Section 4.
%__________________________________________________________________

\section{Analysis and Results}

The stars studied here are the same than those analysed in \citet{abi10} and
\citet{abi11} corresponding to Galactic and extragalactic AGB carbon star samples, 
respectively (thereafter Papers I and II). The extragalactic stars belong to the Carina
dwarf galaxy (2 stars) and the Magellanic Clouds (2 stars in the SMC and 1 star in the LMC).
Details about the observations, data reduction procedures and quality of
the final spectra can be found in these previously published works. The re-determination of fluorine abundances
in these stars has been made using the spectral synthesis method in LTE with the new
version (v14.1) of the $Turbospectrum$ code. Theoretical spectra were computed using the same atmospheric 
parameters (T$_{eff}$, log g, [Fe/H]\footnote{We  use  the
abundance  ratio   definition  [X/Y]$=$  log   (X/Y)$_{\star}-$  log
(X/Y)$_{\odot}$.}, and microturbulence) as in Papers I \& II 
(see those for details), the only difference being the molecular line lists used here (however see below
about changes in the macroturbulence parameter). As noted before, 
the new HF line excitation energies from \citet{jon14b} are now consistent with the partition functions adopted 
in the $Turbospectrum$ code. The immediate consequence of the systematically 0.25 eV lower excitation energies is
that, in AGB carbon stars most of the HF lines in the 2.3 $\mu$m region are saturated 
(log ($W_\lambda/\lambda)\geq -4.0$), even for moderate (Solar)
fluorine abundances. This makes the HF lines more sensitive to the microturbulence
parameter {\bf (see Figure 1)} and, as a consequence, the total uncertainty in the derived F abundances would be larger.
{\bf For instance a change in the microturbulence by $\pm 0.2$ kms$^{-1}$ may change the F abundance derived up to $\mp 0.15$ dex.}

Updates in the molecular line lists in the $2.3~\mu$m region 
has been made. In particular, we have included the contribution from the HCN molecule 
(H$^{12}$CN/H$^{13}$CN) according to the computations by \citet{har03}. We corrected also 
the wavelengths and intensities of some H$_2$O and OH lines in this spectral region following
the new HITRAN database release. Nevertheless, due to the carbon-rich nature of our stars,
the contribution to the global spectrum of these two later molecules is minimal. This is not the case, however, for
the HCN molecule, which has some contribution by introducing a global extra-absorption ($veil$) that might
diminish the spectral continuum up to $\sim 2-3\%$. Also, a significant change with respect to
the line lists used in previous works is the updated CN line list in this spectral region. Details
on how this CN list was computed can be found in \citet{hed13}. The new CN line list 
has an impact on the global fitting of the observed spectra: theoretical fits are improved 
for most of the stars. For several stars, the C/O ratio corresponding to the best fit to the observed spectrum
needed to be revised when compared to the derived value in Papers I \& II (compare Table 1 in these works with Table 2 here). 
Also, for the stars R Lep and SS Vir in Paper I we could not obtain a good fit to their spectra 
and, thus, were discarded from the analysis. On the contrary, we find now a good global fit to the spectrum of the extragalactic 
carbon star LMC TRM88 and determine its F abundance (previously we had just derived an upper limit F abundance for this star, 
see Table 1 in PaperII). 
The new CN line list also affects the HF lines in different ways: while the R9 line still 
remains almost unblended, the R22, R17 and R16 HF lines, which were used
in \citet{abi10} (see their Table 2), are now  quite blended with CN features {\bf (see Figure 1)} and have to be
considered as secondary F abundance indicators. On the contrary, the R23, R15 and R14 HF lines 
show less blending than before with CN lines and complement the R9 line as
main fluorine indicators.

{\bf Figure 1 shows examples of synthetic fits compared to observed spectra of the stars TX Psc and UU Aur. It can be 
clearly appreciated 
the impact of the new excitation energies of the HF lines, micro- and macroturbulence parameters and that of
the new CN line list used in the 2.3 $\mu$m region on the derivation of the F abundance.}

%
%                                                One column figure
%----------------------------------------------------------- S_vib
   \begin{figure*}
   \centering
   \includegraphics[width=15cm]{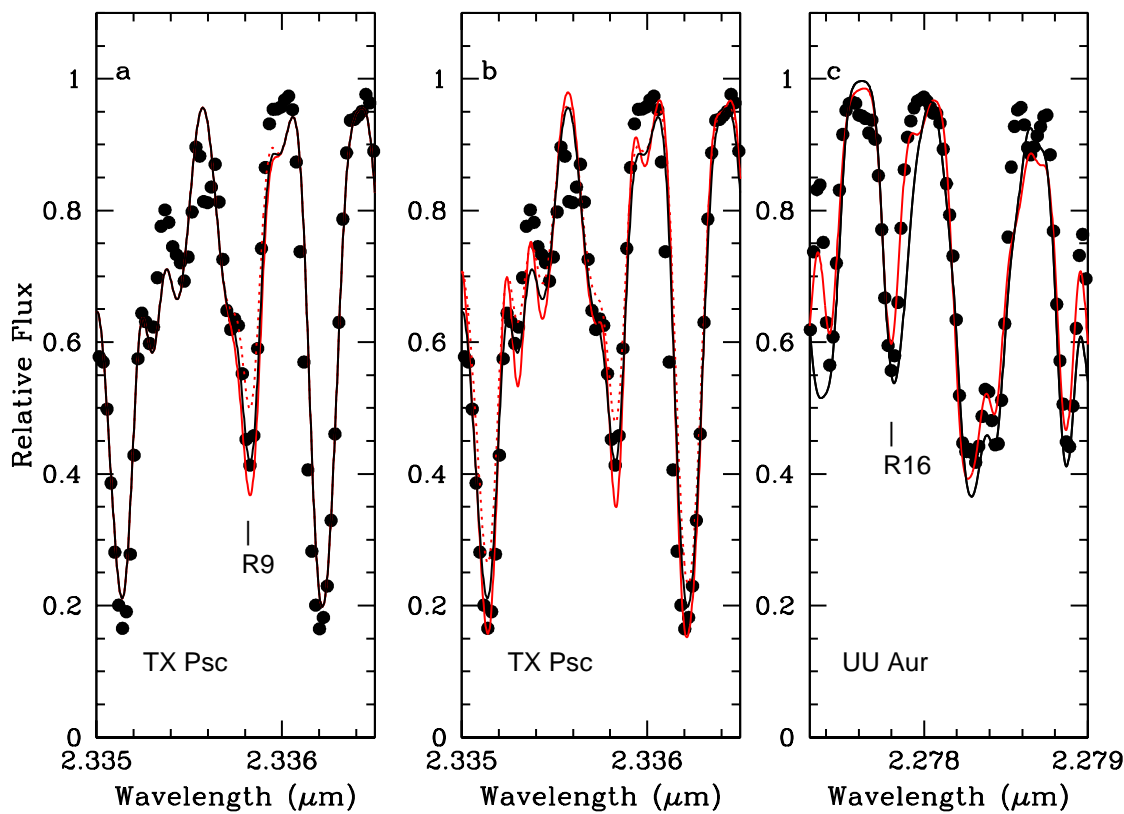}
      \caption{{\bf Examples of synthetic fits (continuous and dotted lines) to the observed spectra (dots) for typical AGB carbon stars in the
region of some HF lines: a) Left. Black line
is the best fit to the R9 HF line in TX Psc (log $\epsilon$(F)$=4.7, \xi=2.2$ kms$^{-1}$ and $\Gamma=13$ kms$^{-1}$); red line is 
a fit assuming a 0.45 dex larger F abundance to show the effect of a F abundance variation similar to the total error 
estimated on [F/H] (see text), while dotted red line shows the corresponding synthetic spectrum adopting the previously wrong 
lower excitation energy for this 
line, $\chi=0.48$ eV. b) Center. Black line the same as case a; red line is a synthetic spectrum adopting a lower macroturbulence 
parameter $\Gamma=11$ kms$^{-1}$, and dotted red line adopting a lower microturbulence, $\xi=1.8$ kms$^{-1}$ than in case a, respectively. 
c) Right. Effect of the new CN line list used in this work (black line) compared with the older one (red line) on the fit to the 
R16 HF line at $\lambda\sim 2.2778$ $\mu$m in the carbon star UU Aur.}}

%         \label{}
   \end{figure*}
%
%______________________________________________________________

\subsection{The Fluorine Abundance in the Sun and Arcturus}
Solar abundances are used as a zero-point for all the abundance
studies. Also, Arcturus is a benchmark star in abundance analyses of
cool giants. To make this analysis as consistent as possible, we
redetermined the F abundance in these two reference stars using the above new molecular parameters
and line lists. For the Sun, we used the sunspot Umbra 
Spectra in the region 1.16 to 2.5 $\mu$m by \citet{wal01}, and followed the same procedure
as in \citet{mai14} to estimate the effective temperature of the atmospheric model
that is most compatible with the sunspot umbra spectrum. We used weak and moderate intensity OH lines 
at 1.5 $\mu$m together with some CO lines at 2.3 $\mu$m. Atomic lines of the metals present in these spectral 
regions cannot be used for this purpose
since most of them are affected by Zeeman splitting. This is not the case for the CO and OH molecular lines, 
nor for the HF lines, because the corresponding vibro-rotational transitions have no spin nor orbital momentum, 
which means a very weak coupling between the rotational momentum and magnetic field. 
In fact, by using a simple Milne atmosphere model and a vertical magnetic field, we checked that 
these molecular lines would be affected by Zeeman splitting only for magnetic fields 
larger than $\sim 10^{6}$ G, an intensity much larger than that observed in the sunspots. 
We estimated a value T$_{eff}=3800$ K from the best fit to the CO and OH lines in the sunspot 
spectrum. Thus we used a MARCS model atmosphere 
with parameters T$_{eff}/$log g/[Fe/H]$=3800/4.44/0.0$ interpolated from a grid of 
atmosphere models \citep{gus08}\footnote{{\bf We checked this approach by using a more realistic 
model atmosphere \citep{col94} for a sunspot which consider the effect of magnetic pressure. 
This sunspot umbra model atmosphere is obtained by analyzing Stokes I and V of several 
spectral lines. With the coolest model atmosphere (T$\sim 3900$ K, B$=2750$ G at 
log $\tau_{5000}=0.0$) deduced by these authors (see their Table 4), which resembles the MARCS model
adopted here, we obtain a similar fit to the CO and OH lines and also a very 
similar average F abundance, log $\epsilon$(F)$=4.38\pm 0.04$ (see above). This implies that magnetic
pressure in a typical sunspot does not affect significantly the derivation of the F abundance.}}. 
Similarly to \citet{mai14} we found that the cores of the stronger CO 
and of some OH lines were not so well matched. This difficulty has already been reported in the analysis
of the infrared spectra of many K giants with T$_{eff}$ similar to that we estimated in the sunspot 
\citep[e.g.][]{ryd02,tsu08,tsu09} suggesting that 
the CO fundamental lines cannot be interpreted with a photospheric model only. These lines show an 
excess absorption (log W$_\lambda/\lambda\geq$ -4.75) that is probably 
nonphotospheric in origin, indicating that the structure of the hydrostatic 
atmosphere does not describe properly that of the sunspots\footnote{Nevertheless, it is possible to fit
reasonably well the cores of these molecular lines by extending the atmosphere model up to log $\tau_{5000}\sim -6$.
This extension of the photosphere does not affect at all the HF lines, which indicates that 
these lines form into deeper 
layers where the structure of our model atmosphere for the sunspot would be more realistic \citep[see e.g.]
[for alternative solutions]{tsu09}.}. \citet{mai14} arrived at the 
same conclusion, nevertheless they estimated a T$_{eff}=4250$ K as the best value matching 
both the OH and CO lines in the sunspot spectrum. Nevertheless, we derive 
log $\epsilon$(F)$=4.37\pm 0.04$ in the Sun from 10 HF lines, in agreement 
with the value estimated by \citet{mai14} and with the meteoritic abundance \citep{lod09}. 
The uncertainty on this value is mainly determined by the error in the temperature estimate of 
the sunspot spectrum. \citet{mai14} quoted about 0.25 dex for this error that we also
adopt here. In the following we will use our derived F abundance as the reference value for the Sun.

For Arcturus, we used the electronic version of the Infrared Atlas
Spectrum by \citet{hin95} and the atmosphere parameters from \citet{ryd09}. We obtain a fluorine abundance of 
log $\epsilon$(F)$=3.78\pm0.03$
from the analysis of the R7, R9 and R12 HF lines. This value is also in excellent agreement with that
obtained by \citet{jon14a} and \citet{jon14b}, the later from the analysis of some HF lines at 12.2 $\mu$m.

\subsection{Fluorine Abundances in AGB Carbon Stars}

Table 2 (second column) shows the fluorine abundances redetermined in the sample of Galactic and
extragalactic AGB carbon stars of different spectral types.  For each star
we report the number of HF lines used (between parenthesis) and the corresponding dispersion 
when more than a single line was used. Table 2 also shows (third column) the difference in the F abundance derived 
with respect to those in Papers I \& II. The mean difference is $-0.33\pm 0.17$ dex, i.e. the abundances 
derived here are on average systematically lower by this amount. Note that for many stars 
the reduction in the F abundance is different from that applying just the correction factor 
$\sim 5040/T_{eff}\Delta\chi$ (see Table 1). The reason is twofold: a) the impact of the new CN line list and, b) 
the use here of larger macroturbulence velocities ($\Gamma$) with respect to those used in Papers 
I $\&$ II. The new CN line list and partition functions affects the HF lines in a different manner depending on the 
actual C/O ratio and the metallicity of the star (see Table 1). The {\bf former} effect 
depends on the blending of a specific HF line with CN lines {\bf(see Figure 1)} being this, in any case, no larger 
than {\bf $\sim\pm 0.20$ dex}. On the other hand, because the use of new molecular line lists (mainly CN), 
we realised that the 2.3 $\mu$m spectral
region is better fitted using a macroturbulence parameter in the range $13-14$ kms$^{-1}$ {\bf(see Figure 1)}, 
instead of $11-12$ kms$^{-1}$ adopted in 
Papers I \& II. This might increase the fluorine abundance
as much as 0.25 dex, compensating partially the systematic decrease of the F abundances because of the lower $\chi$ values. 
In summary, the F abundances derived here differ from the previous values by an amount which is different star by
star depending on the combined effect of the lower $\chi$ values, the higher $\Gamma$ parameters and  
the blending with CN lines, {\bf in this order of relevance.} 
   
The total uncertainty in the abundance determination can be determined from 
the dependence of the fluorine abundance on the stellar atmospheric parameters; this dependence
is rather similar for all the HF lines, being the uncertainty in T$_{eff}$, $\xi$, $\Gamma$ and 
the C/O ratio the main sources of error. {\bf A detailed discussion on the error sources and their 
impact in the final F abundance can be found in \citet{abi09}. Also a discussion on the uncertainties
in the derivation of the stellar metallicity [Fe/H] and $s$-element abundances can be found in
\citet{abi01,abi02,deL06} and \citet{abi08}. Concerning F,} the quadratic addition of all these sources of error gives a 
typical uncertainty of $\pm 0.40$ dex, together with the uncertainty of the continuum position and the dispersion 
in the F abundance when derived from several lines, a conservative total error would 
be $\pm 0.45$ dex. This value does not include possible systematic errors as those as in 
the model atmospheres structure and/or N-LTE effects. However, the uncertainty in the abundance ratio between F and any 
other element would be lower than this value, since some of the above uncertainties cancel 
out when deriving the abundance ratio. Note, however, that the new [F/Fe] ratios in many stars
(column 5 in Table 2), do not differ significantly from the previous ones because the 
reduction of the solar F abundance by 0.19 dex (see previous section).

%
%                                                One column figure
%----------------------------------------------------------- S_vib
   \begin{figure}
   \centering
   \includegraphics[width=8cm]{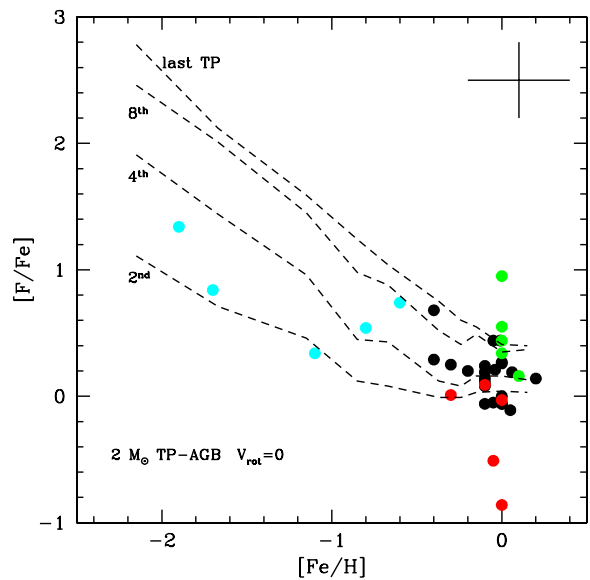}
      \caption{[F/Fe] vs. [Fe/H] in the stars analysed. Circles data points 
are: black, N-type Galactic carbon stars; red, J-type; green, SC-type and blue, extragalactic N-type carbon stars. 
Dashed lines represent theoretical predictions for a non-rotating 2 M$_\odot$ TP-AGB star 
at different TPs according to \citet{cri15}. A typical error bar
is indicated.}
%         \label{}
   \end{figure}
%
%______________________________________________________________

\section{Discussion}

The main consequence of the redetermined F abundances is that 
they are systematically lower than those in previous determinations, in particular for carbon stars of near solar
metallicity. As it can be seen in Table 2, normal (N-type) carbon stars show only
moderate F enhancements ([F/Fe]); the largest enhancements are now close to $\sim 1$ dex 
but are observed only in one SC-type star (WZ Cas)  and 
some of the metal-poor N-type stars in Carina and the Magellanic Clouds. On the other hand, J-type carbon stars show 
no F enhancements or, in some cases, even depletions (RY Dra and Y Cvn, see Table 2). We confirm, 
therefore, the results of Paper I concerning the [F/Fe] ratios in AGB carbon stars of different spectral types. 
These new enhancements can still  be accounted for by current 
low-mass TP-AGB nucleosynthesis models {\bf (see below).}

%----------------------------------------------------------- S_vib
   \begin{figure}
   \centering
   \includegraphics[width=8cm]{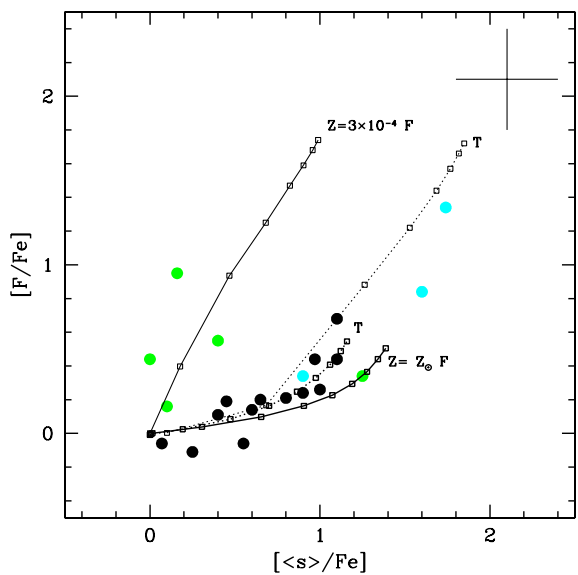}
      \caption{Observed fluorine vs. average s-element enhancements compared 
with theoretical predictions. Circles symbols are as in Figure 2. Solid lines are the predictions for 
a non-rotating 2 M$_\odot$ TP-AGB with metallicities (labelled) mimiking those of our stellar
sample according assuming the standard $^{13}$C-pocket according to \citet{cri15} (labelled with {\it F}, FRUITY). 
Dotted lines is the prediction for a non-rotating 1.3 M$_\odot$ metal-poor model for the same metallicities in 
the hypothesis of a larger $^{13}$C-pocket (labelled with {\it T}, Tail) (see text). Small open squares in these lines mark the values 
predicted at different TPs. J-type AGB carbon stars have been excluded because 
they usually do not show s-element enhancements \citep{abi00}. Typical uncertainties 
are shown. See text for details. }
%         \label{}
   \end{figure}
%
%______________________________________________________________
We compare the revised F abundances with theoretical models computed with the FUNS evolutionary 
code \citep{stra06,cri09}. Those models are available on line on the web pages of the 
FRUITY\footnote{http://fruity.oa-teramo.inaf.it.} database,
which includes stars with initial masses 1.3 $\le$ M/M$_\odot \le$ 6.0 and metallicities -2.15 $\le$ [Fe/H] $\le$ +0.15
\citep{cri11,cri15}. Those models are calculated by coupling the physical evolution of the 
structure with a nuclear network including all chemical species, from hydrogen up to Pb-Bi 
(at the termination of the s-process path). Thus, no post process techniques are used.
In FRUITY models, a thin $^{13}$C pocket forms after each TDU episode. The mass extension of 
the pockets decreases along the AGB, thus making the first pockets the most significant for 
the on-going s-process nucleosynthesis. The mass-loss rate is calibrated 
on the observed mass loss - period relation found in Galactic AGB stars \citep[see][and references therein]{stra06}. 
In order to properly follow the physical behaviour of the most external layers, 
low temperature C and N enhanced molecular opacities are used \citep{cri07}. In fact, if the convective 
envelope becomes C-rich (i.e. C/O$>$1), C-bearing molecules form largely contributing to the 
opacity (more than O-bearing molecules). Thus, the structure becomes more opaque to photons 
and the external layers expand and cool. This implies an increased mass-loss rate. Note that, 
with respect to previously published FRUITY low mass AGB star yields \citep{cri11}, we find lower F surface abundances due 
to a revision in the opacity treatment of the sub-atmospheric region \citep{cri15}. 
{\bf We confirm previous calculations \citep{lug04,kar10} showing that the stellar mass range for F production
peaks around 1.5-2.5 M$_\odot$ for all Z, showing a small dependence on the metallicity.
This implies that very large surface [F/Fe] ratios are predicted for decreasing metallicities.
Net fluorine yields from IMS stars are lower than their low mass counterparts, but indeed have a stronger
dependence on the metallicity. In fact, in our low Z IMS models the $^{19}$F$(p,\alpha)^{16}$O and $^{19}$F$(\alpha,p)^{22}$Ne 
reactions are more efficiently activated, both leading to a net fluorine destruction. In any case, however, the IMS 
contribution to the fluorine is negligible, due to the fact that these stars experience 
a definitely less efficient TDU \citep[see][]{cri15}.}

Figure 2 shows the [F/Fe] ratios derived in our stars versus the stellar metallicity. The different dots
refer to the different spectral types carbon stars in the Galaxy (see figure caption).
Blue dots are the extragalactic metal-poor carbon stars belonging to the satellite galaxies. 
From this figure 
it is evident that 
J-type stars tend to be F-poor objects compared to the normal N-type (black dots) carbon stars. Moreover, 
there is a hint that SC-type stars show, on average, larger F enhancements with respect to N-type stars. With
the present analysis we confirm what we found in Paper I in this respect. Note that the origin 
of SC- and J-type stars is still unknown. There is evidence that they are outside the classical 
spectral evolution M-MS-S-C(N) along the AGB phase, which is thought to be a consequence
of the progressive carbon enrichment of the stellar envelope because of TDU episodes\footnote{For a detailed
discussion of the properties and possible evolutionary status of SC- and J-type carbon stars see \citet{wal98} and
\citet{abi03}.}. Nevertheless, the small number of J- and SC-type stars studied prevent us to reach a 
definite conclusion. We notice from Fig. 1 that Galactic N-type stars with slightly lower [Fe/H] present 
larger [F/Fe]. This trend is confirmed for the extragalactic AGB carbon stars (blue dots), which show definitely larger fluorine 
overabundances, thus confirming  the results of Paper II. Dashed lines in Figure 2 are the predicted [F/Fe] 
ratios as a function of the metallicity for a typical non-rotating 
2 M$_\odot$ AGB star at different TPs (as labelled). It can be seen that observed 
and predicted [F/Fe] ratios agree well within observational errors. Furthermore, the predicted
C/O ratios in the envelope at the observed metallicity of the stars agree with the
corresponding observed F enhancement. For instance, in the {\bf FRUITY database} models with [Fe/H]$\sim 0.0, -1.1, -2.1$ 
(or Z$=0.014, 10^{-3}$ and 10$^{-4}$, respectively) at the 3$^{rd}$ TP, the 
predicted ratios are C/O$\sim 1.1, 2.0$ and 15, respectively. These C/O ratios are compatible with
those measured in our sample at those metallicities (see Table 2). In brief, we confirm here
the predicted dependence of the F 
production with the metallicity \citep[e.g.][]{lug04,cri11,fis14}.

A correlation between the F and the $s-$element overabundance is expected 
if a large enough $^{13}$C pocket forms after each TDU episode. This correlation
comes from the $^{15}$N production in the radiative $^{13}$C pocket, the site where the s-process main 
component is built up (see Paper I for details). Figure 2 shows the new [F/Fe] ratios vs. the observed average s-element enhancement
(Abia et al. 2002; de Laverny et al. 2006; Abia et al. 2008). For the solar-like metallicity
N-type carbon stars (black dots), F and s-element overabundances clearly correlate. Such a
trend seems to hold for metal-poor extragalactic carbon stars (blue dots) as well, with the largest
fluorine overabundances corresponding to the most s-process rich stars.
In Figure 3 we compare observations with FRUITY models having two different (M,Z) combinations, i.e. 
initial mass M$=2$ M$_\odot$ with [Fe/H]$=0$ and M$=1.3$ M$_\odot$ with [Fe/H]$=-1.67$.
Both curves start from FDU abundances (solid curves). The two metalliciteis are representative
of galactic and extragalactic carbon stars, respectively. At low Z we choose a lower initial mass in order 
to be consistent with the bolometric magnitudes
of the observed stars ($-4.3 < $M$_{bol} < -5.3$). At solar metallicity observations {\bf(excluding most of SC stars)}
and theory {\bf(the two lower curves in Figure 3)} agree nicely. At low Z {\bf(upper continuos line in Fig.3)}, 
instead, we highlight a clear disagreement, 
the models showing too 
large $^{19}$F abundances for a fixed s-process enrichment
(see the discussion in Abia et al. 2011). A similar apparent deficiency of F abundances for a given s-element enhancement
has been found also in the Galactic Carbon Enhanced Metal-Poor stars (CEMP) which are believed to have been polluted by mass-transfer
from a former AGB star in a binary system (e.g. Lucatello et al. 2011). 
We test whether the inclusion of rotation
might alleviate this problem by modifying the  $^{19}$F surface distributions in our models (see
Piersanti et al. 2013). The main fluorine production channel starts with the neutrons released by the
$^{13}$C$(\alpha,n)^{16}$O reaction and proceeds through the nuclear chain $^{14}$N$(n,p)^{14}$C$(\alpha,\gamma)^{18}$O$(p,\alpha)^{15}$N
$(\alpha,\gamma)^{19}$F \citep[see][and references therein]{cri14}. The main effect of mixing induced by rotation is
to dilute $^{14}$N within the $^{13}$C-pocket. Thus, more neutrons are captured by $^{14}$N, feeding the nuclear
path leading to $^{19}$F production. This has also important consequences on the on-going s-process
nucleosynthesis, which results less efficient depending on the initial rotation velocity ($^{14}$N is in fact
the major neutron poison of the s-process). Thus, rotation leads to slightly larger fluorine surface
abundances and to a definitely lower s-process efficiency.
We conclude, therefore, that this physical process cannot be considered as a potential candidate to solve the fluorine 
discrepancy between theory and observations at low metallicities.
A possible way to solve such a discrepancy is to obtain larger s-process enhancements for a fixed dredged up mass.
\citet{cri15} recently demonstrated that a different treatment of the inner boundary of the extra-mixed region at 
the base of the convective envelope during a TDU may lead to appreciable differences in the surface s-process distribution. 
In particular, those authors found that allowing the partial mixing to work below the formal Schwarzschild convective boundary down to the
layer where the convective velocity is 10$^{-11}$ times lower, a net increase of s-process abundances can be obtained. This derives 
from the fact that $^{13}$C pockets are larger than those obtained in standard FRUITY models (in which the lower 
boundary is fixed at 2 pressure scale heights from the formal
border of the convective envelope). In Figure 3 we report two models with the same combination of mass 
and metallicity as the FRUITY models already discussed, but with the aforedescribed different mixing treatment 
at the base of the convective envelope ({\bf dotted} lines in Fig. 3). At solar-like metallicity we find a slight decrease in the fluorine 
surface overabundances, this fact however not compromising the agreeement with observations. At low Z, instead, we notice a large 
fluorine reduction for a fixed s-process surface enhancement. This leads to a reasonable fit to extra-galactic carbon stars.
Thus, we conclude that larger $^{13}$C pockets than those characterizing FRUITY models are needed to fit 
fluorine abundances in the metal-poor extra-galactic carbon stars. This further strenghten similar conclusions already 
reached in the study 
of s-process rich stars belonging to open clusters \citep{maiorca,trippa} as well as in the isotopic analysis of pre-solar SiC 
grains \citep{liu14,liu15}. Note that, in any case, SC-type carbon stars marginally agree with theoretical models. 
However, the evolutionary status of these objects is not well understood.

%----------------------------------------------------------- S_vib
   \begin{figure}
   \centering
   \includegraphics[width=8cm]{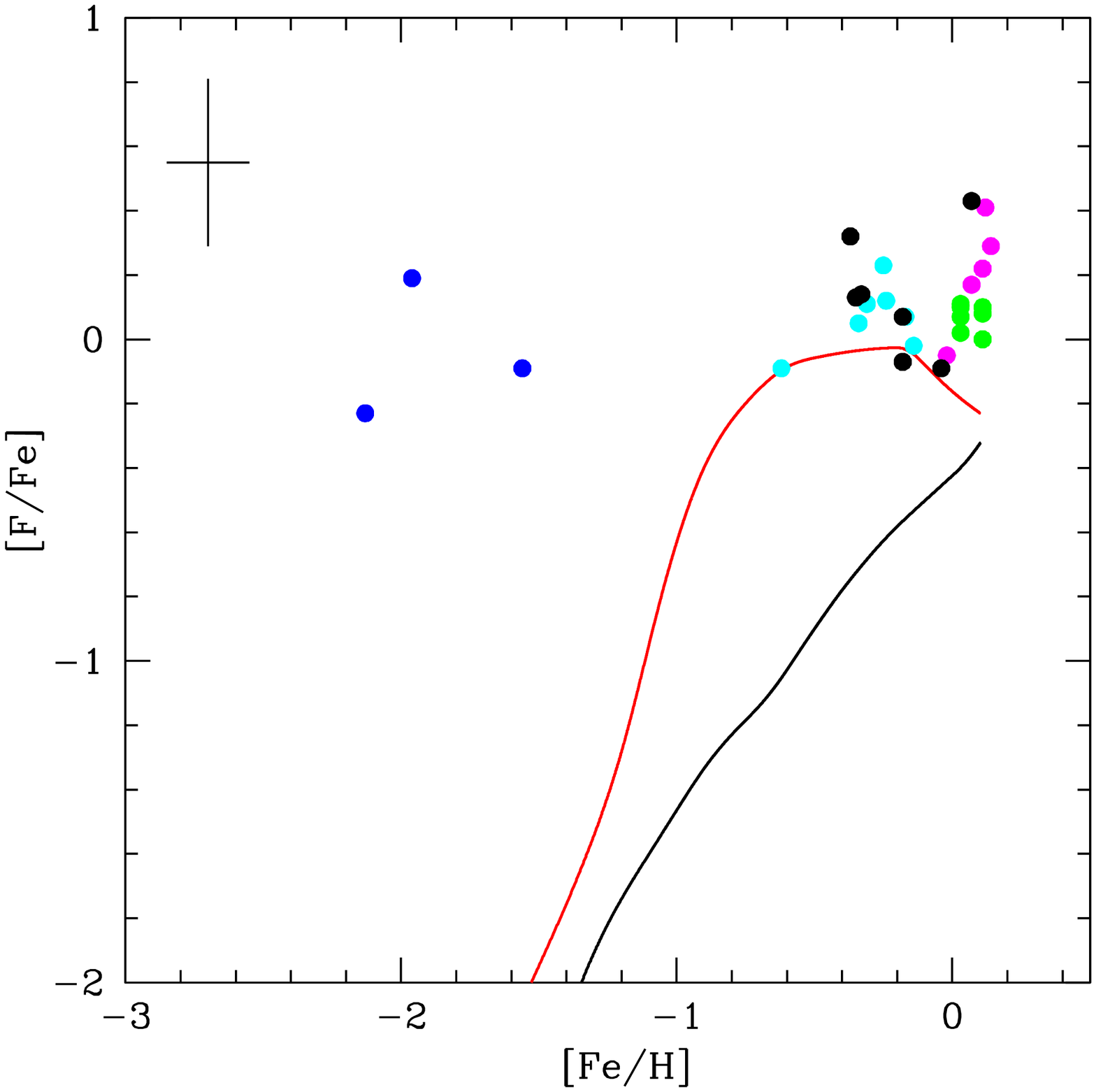}
      \caption{[F/Fe] vs. [Fe/H] observed in Galactic field 
dwarfs and giants and open cluster stars compared with the predicted evolution from 
our GCE model for the solar neighbourhood including only fluorine production from AGB stars. 
Black line: yields from \citet{cri15b}; Red line: yields from \citet{kar10}. 
The observed ratios (circles) are from \citet{ale12} (black), \citet{li13} (blue), \citet{mai14} (green), 
\citet{jon14b} (cyan) and \citet{nau13} (magenta). A typical error bar 
is shown. See text for details.}
%         \label{}
   \end{figure}
\subsection{The contribution of AGB stars to the fluorine abundance in the Solar neighbourhood}
As mentioned in Section 1, the real contribution of AGB stars to the F inventory in the Solar neighbourhood is still a 
matter of debate \citep{jon14a}. The present downward revision of the F overabundances 
observed in AGB stars and new model predictions re-opens this issue. 
To shed light on this problem, we have computed a simple Galactic Chemical Evolution (GCE) model 
\citep{cri15b} used recently to study the Solar System s-only distribution (i.e. of those isotopes uniquely synthesised by
AGB stars). This GCE model accounts for all the constraints 
currently observed in the Solar neighbourhood and at the epoch of the Solar System formation \citep{boi99}. By using this
code we have calculated the [F/Fe] vs. [Fe/H] evolution including only the F contribution from AGB stars. Figure 4
shows the computed evolution (black line) compared to the observed one inferred from Galactic unevolved field dwarf 
and giant stars {\bf and open cluster stars} (see caption of Figure 4 for the references\footnote{{\bf All the [F/Fe] ratios shown 
in Figure 4 have been re-scaled to the new solar F abundance according to Table 2. Note that these ratios
may not differ significantly from the original ones in the mentioned papers, since apart of the lower Solar
F abundance used here, one should also consider the decrease of the F abundance derived in a specific star
when using the correct $\chi$ values.}}). It is evident that the AGB contribution to F is not 
enough to account for the observed abundance 
of this element in the Solar System: the GCE model predicts a factor $\sim 3$ lower [F/Fe] ratio than observed at
[Fe/H]$\sim0.0$. We have computed also the evolution using the F yields obtained with the STROMLO 
stellar evolutionary code \citep[see e.g.][red line]{kar10}. From Figure 4, it is evident 
that the GCE model produces quite different results when using different $^{19}$F yield inputs. The reasons for such a discrepancy 
are manifold. Among the many different inputs of the two stellar codes, two main candidates can be identified: 
the treatment of convection and the mass-loss rate. With respect to ours models, those of \citet{kar10}
show an increased TDU and hot bottom burning efficiency. While the first difference leads to 
larger $^{19}$F surface enrichments, the second implies a more efficient fluorine destruction. 
However, when weighting the contribution with a typical initial mass function, the first term becomes 
dominant. As a consequence, the fluorine production in Karakas's stellar models is larger than our.
This is particularly evident for stellar models with initail mass M$\ge 3$ M$_\odot$ at low metalliciy (Z$\leq 6\times 10^{-3}$).  
Another difference arises from the adopted mass-loss rate law. The models by \citet{kar10} follow the 
prescriptions of \citet{vw93}. At fixed period, our mass-loss rate is more efficient 
(see Figure 6 in Straniero et al. 2006). This leads to a faster erosion of the convective 
envelope and, thus, to lower yields. Moreover, \citet{kar10} included a metallicity 
dependency that we do not take into account. In any case, using the Karakas (2010) yields for F (red line), the computed evolution 
agrees better with the observed trend at 
[Fe/H]$\sim 0.0$, but the previous conclusion still holds: AGB stars alone cannot account for the current F abundance 
in the Solar neighbourhood and other sources of F are required to explain its observed evolution. {\bf Note that although using
the Karakas's yields one might account for the observed [F/Fe] ratios at [Fe/H]$\sim -0.5$ (within the
observational uncertainties, see Fig. 3),  at
higher metallcities her yields fail to get [F/Fe]$\sim 0.0$ as the observations seems to indicate on average.}
Additional fluorine abundance determinations in dwarf and giant stars covering a wide range of metallicities 
would help to clarify the nature of these additional sources.
We notice, furthermore, that current stellar theoretical models reproduce the $^{19}$F abundances observed in AGB stars. 
This further strengthen the need of additional fluorine sources.

\section{Summary}

  Fluorine abundances have been redetermined in a sample of Galactic and extragalactic AGB carbon stars using consistent partition
functions and spectroscopic parameters of HF lines. A redetermination of the F abundance in the Sun and Arcturus has been also made, 
which agree with other recent F abundance determinations in these stars. Concerning AGB carbon stars, the new F abundances are systematically 
lower in average by 0.33 dex with respect to previous determinations. For a specific carbon star the difference depends on a 
combination between the new excitation energies of the HF lines, a higher macroturbulence parameter, and a new CN line list used here. 
The F abundances found in the near solar metallicity carbon stars agree with the extant nucleosynthesis models in low-mass AGB stars. 
For the metal-poor AGB carbon stars belonging to several satellite galaxies, a satisfactory fit can be obtained by using theoretical 
models with larger $^{13}$C pockets with respect to our standard FRUITY models. Using new F yields from low and intermediate-mass AGB 
stars and a GCE model we evaluate the contribution of these stars to the F inventory in the solar neighbourhood and 
conclude that additional sources of F are needed to explain the observed evolution of this element.

\begin{acknowledgements}
Part of this work was supported by the Spanish grant
AYA-2011-22460. NSO/Kitt Peak FTS data used here were produced by NSF/NOAO.
{\bf We thank A. Asensio Ramos for his help on the atmosphere models for
sunspots.}
\end{acknowledgements}

%-------------------------------------------------------------------
\bibliographystyle{aa} % style aa.bst
\bibliography{fluor2.bib} % your references Yourfile.bib

\begin{table*}
\caption{Variations (new-old) in the F abundance derived (in dex) from some HF lines due o the combined effect of the
new $\chi$'s and CN molecular list for specific atmosphere parameters}% title of Table
\label{table:1}      % is used to refer this table in the text
\centering                          % used for centering table
\begin{tabular}{ccccccc}        % centered columns (4 columns)
\hline\hline                 % inserts double horizontal lines
         &       &      &     &  & HF lines     & \\
\hline
T$_{eff}$ & log g &[Fe/H]& C/O &   R9 &  R14 & R15\\    % table heading 
\hline                        % inserts single horizontal line
2800&0.0&0.0&1.02                          &-0.40  &-0.35 & -0.35\\
3300&0.0&0.0&1.02                          & -0.40 &-0.35 & -0.35 \\ 
2800&0.0&0.0&1.10                          & -0.40 &-0.35 & -0.40\\ 
2800&0.0&-1.0&1.70                         & -0.45  &-0.45 & -0.45\\ 
4000&3.0&0.0&0.58                          & -0.33 &-0.30 &-0.25\\
5500&4.0&0.0&0.58                          & -0.30 &-0.30& -0.30\\
\hline
\end{tabular}
\tablefoot{It is assumed [F/Fe]$=0.0$, {\bf a macroturbulence} $\Gamma=13$ kms$^{-1}$ in all the cases, and 
{\bf microturbulence} $\xi=2.2$ or 1.5 kms$^{-1}$, for carbon- or oxygen-rich stars, respectively.\\
}
\end{table*}

\begin{table*}
\caption{Fluorine Abundances and Element Ratios}             % title of Table
\label{table:2}      % is used to refer this table in the text
\centering                          % used for centering table
\begin{tabular}{lcccccc}        % centered columns (4 columns)
\hline\hline                 % inserts double horizontal lines
Star & log $\epsilon$(F)\tablefootmark{a}& $\Delta$log $\epsilon$(F)\tablefootmark{b} & C/O & {\bf[Fe/H]}&[F/Fe]&
[$<$s$>$/Fe]\\    % table heading 
\hline                        % inserts single horizontal line
Sun       & 4.37$\pm0.04$(10)&0.19&0.58&0.00&0.00&0.00\\
Arcturus  & 3.78$\pm0.02$(3) &0.37&0.20&$-0.53$&$-0.07$&$-0.01$   \\
          &                  &     &   &     &  \\
N-type   &                  &     &   &     &  \\
          &                  &     &   &     & \\
AQ Sgr& 4.34$\pm0.08$ (5)& 0.29& 1.03& $-0.10$&0.08& --\\
BL Ori& 4.62$\pm0.20$ (5)& 0.30& 1.04& 0.00   & 0.26& 1.00\\
RT Cap& 4.30             & 0.36& 1.20&$0.00$&       $-0.06$& --\\
RV Cyg& 4.30$\pm0.00$ (4)& 0.40& 1.09&0.00&$-0.06$& 0.07\\
S Sct & 4.37$\pm0.05$ (5)& 0.25& 1.05&$-0.10$&        0.11& 0.40\\
ST Cam& 4.45$\pm0.00$ (4)& 0.27& 1.02&$-0.10$&       0.19& 0.45\\
TU Gem& 4.63$\pm0.24$ (5)& 0.29& 1.07&0.00&        0.27& --\\
TW Oph& 4.26$\pm0.06$ (3)& 0.35& 1.12&$-0.40$&        0.29&--\\
TX Psc& 4.65$\pm0.07$ (6)& 0.18& 1.03&$-0.39$&       0.69& 1.10\\
U Cam & 4.20$\pm0.08$ (4)& 0.38& 1.23&$-0.10$&        $-0.06$& 0.55\\
U Hya & 4.70$\pm0.20$ (6)& 0.37& 1.05&$-0.05$&       0.39& 1.10\\
UU Aur& 4.55$\pm0.14$ (7)& 0.33& 1.06&$0.00$&        0.19& 0.45\\
UX Dra& 4.56$\pm0.18$ (5)& 0.29& 1.05&$-0.20$&       0.20& 0.65\\
V460 Cyg&4.47$\pm0.12$ (4)& 0.18& 1.07&$-0.05$&      0.15& 0.80\\
V Aql & 4.26$\pm0.05$ (5)& 0.36& 1.05&$-0.05$&       $-0.05$& 0.20\\
VY UMa& 4.50$\pm0.03$ (5)& 0.20& 1.06&$-0.10$&       0.24& 0.90\\
W CMa & 4.68$\pm0.13$ (5)& 0.27& 1.07&0.20&        0.12& 0.60\\  
W Ori & 4.30$\pm0.00$ (3)& 0.12& 1.07&$0.05$&       $-0.10$& 0.25\\
X Cnc & 4.32$\pm0.08$ (4)& 0.61& 1.03&$-0.30$&        0.25& --\\
Y Hya & 4.40$\pm0.00$ (2)& 0.50& 1.18&$-0.10$&        0.14& --\\
Y Tau & 4.36$\pm0.17$ (4)& 0.21& 1.04& 0.00&       0.00& 0.05\\
Z Psc & 4.80$\pm0.06$ (5)& 0.10& 1.08&$-0.01$&        0.44& 0.97\\
      &                  &     &     &     & \\
SC-type&                &     &     &     & \\
        &                &     &     &     & \\
CY Cyg& 4.62$\pm0.03$ (4)& 0.53& 1.00&0.10&         0.15& 0.10\\
FU Mon& 4.91$\pm0.07$ (5)& 0.24&1.00& 0.00&         0.55& 0.40\\
GP Ori& 4.70$\pm0.14$ (5)& 0.56&1.01& 0.00&         0.34& 1.25\\
RZ Peg& 4.80$\pm0.02$ (7)& 0.20&1.00& 0.00&         0.44& 0.00\\
WZ Cas& 5.31$\pm0.14$ (7)& 0.49&1.01& 0.00&         0.95& 0.16\\
      &                  &     &    &     &\\
J-type&                 &     &    &     &\\
       &                 &     &    &     & \\
R Scl & 4.07$\pm0.05$ (4)& 0.11&1.02 &$-0.30$&        0.01& -- \\
RY Dra& 3.80             & 0.68& 1.12&$-0.05$&         $-0.51$& $<$0.20\\
T Lyr& 4.32              & 0.35&1.14&$0.00$&         $-0.04$& --\\
VX And& 4.30$\pm0.00$ (2)& 0.05&1.76&$-0.10$&        0.04& 0.2\\
Y Cvn&  $<3.50$          & $>0.85$&1.12&0.00& $<-0.86$& $<$0.2\\
     &                   &     &    &  &\\
ExtraGalactic N-type &  &     &    &  &\\
                      &  &     &    &  & \\
LMC TRM88& 4.50  &0.35&1.62  & $-0.60$&          0.74& --\\
SMC GM780& 4.10  &--&3.10    & $-0.80$&           0.53& --\\
SMC BMB B30&3.60 &0.00&1.47  & $-1.10$&           0.34& 0.90\\
Carina ALW6&3.50 &0.20&12.0  & $-1.70$&           0.84& 1.60\\
Carina ALW7&3.80 &0.60&19.0  & $-1.90$&           1.34&1.74\\
\hline                                   %inserts single line
\end{tabular}
\tablefoot{[$<$s$>$/Fe] is the average s-element enhancement according to \citet{abi00}, \citet{abi02}, \citet{deL06} and \citet{abi08}.\\
\tablefoottext{a}{The number in parenthesis indicates the number of HF lines used.}
\tablefoottext{b}{Abundance difference in the sense \citet{abi10} (Galactic stars) or
\citet{abi11} (extragalactic stars) minus this work. In the
cases of the Sun and Arcturus the differences are respect to that derived in \citet{aps09} 
and \cite{abi09}, respectively. }
}
\end{table*}
\end{document}